# Insight into $CO_2$ Dissociation in Plasmas from Numerical Solution of a Vibrational Diffusion Equation

*Paola Diomede[1,*], Mauritius C. M. van de Sanden[1], Savino Longo[2]*

[1] *DIFFER – Dutch Institute for Fundamental Energy Research, P.O. Box 6336, 5600 HH Eindhoven, The Netherlands*

[2] *Dipartimento di Chimica, Universita' degli Studi di Bari, via Orabona 4, 70126 Bari, Italy*

*Corresponding Author

E-mail: p.diomede@differ.nl

Tel: +31-(0)40-3334-924

**Abstract**

The dissociation of $CO_2$ molecules in plasmas is a subject of enormous importance for fundamental studies and in view of the recent interest in carbon capture and carbon-neutral fuels. The vibrational excitation of the $CO_2$ molecule plays an important role in the process. The complexity of the present state-to-state (STS) models makes it difficult to find out the key parameters. In this paper we propose as an alternative a numerical method based on the diffusion formalism developed in the past for analytical studies. The non-linear Fokker-Planck equation is solved by the time-dependent diffusion Monte Carlo method. Transport quantities are calculated from STS rate coefficients. The asymmetric stretching mode of $CO_2$ is used as a test case. We show that the method reproduces the STS results or a Treanor distribution depending on the choice of the boundary conditions. A positive drift, whose energy onset is determined by the vibrational to translational temperature ratio, brings molecules from mid-energy range to dissociation. Vibrational-translational energy transfers have negligible effect at the gas temperature considered in this study. The possibility of describing the dissociation kinetics as a transport process provides insight towards the goal of achieving efficient $CO_2$ conversion.

**Introduction**

Plasma-assisted gas conversion techniques are widely considered as efficient building blocks in a future energy infrastructure which will be based on renewable but intermittent electricity sources. In particular $CO_2$ dissociation in high-frequency plasmas is of interest in carbon capture and utilization process chains for the production of $CO_2$-neutral fuels[1]. In this case, the vibrational excitation of the $CO_2$ molecule plays an important role in the energy efficient non-equilibrium dissociation kinetics, however several aspects of the dissociation kinetics in plasmas are still unclear.

Dissociation takes place when collisions between molecules and electrons, as well as inter-molecular collisions, provide enough energy to lead an already excited molecule into the continuum region thereby producing CO and O. The state-to-state (STS) approach[2,3] allows to calculate very accurate reaction rates by considering any vibrational state as an individual species. This amounts to solve the so-called Master Equation (ME) for the populations of vibrational states[2,4]: The ME is actually a system of $n$ non-linear ODEs (Ordinary Differential Equations) where *n* is the number of vibrational states considered, with a complex right hand side including terms for any possible chemical process, each in the form of a product of the values of the vibrational distribution function (VDF) and a temperature dependent rate coefficient.

The rate coefficients are selected from literature data or, for electron induced processes, calculated from the related cross sections and the electron energy distribution function (EEDF). The main problem in calculating the VDF for a polyatomic molecule is the large number of states leading to a huge number of possible transitions between them. This leads to high computational costs and requires large sets of data, most of which not well known.

Specially in the past few years, strong efforts have been devoted to reducing the complexity of the kinetics of $CO_2$ dissociation in plasmas, after many years of inactivity after the works of Gordiets and other scientists on lasers kinetics.[4,5] The current availability of computational resources allows the implementation of multi-dimensional fluid models for the description of the non-thermal plasmas where $CO_2$ is activated, but still, the complexity of the plasma/neutral chemistry has to be coupled to the flow of the gas/plasma and to the electromagnetic field used to generate the plasma. A further complication is given by the presence of multiple time and length scales in the kinetics and dynamics of neutral and charged particles.

In the case of $CO_2$ molecules, three vibrational modes have to be accounted for: symmetric stretching, doubly degenerate bending and asymmetric stretching. Usually the formalism $CO_2(i,j,k)$ is used where i,j,k are respectively the numbers of quanta in these modes. The redistribution of internal energy in $CO_2$ is the result of a series of elementary processes, including VT (vibration to translation) energy exchanges $CO_2(i,j,k) + X \rightarrow CO_2(u,w,v) + X$ where internal energy is partially converted into kinetic energy and VV (vibration to vibration) energy exchanges, $CO_2(i,j,k) + CO_2(l,m,n) \rightarrow CO_2(u,w,v) + CO_2(a,b,c)$, where most of the internal energy is kept as such but redistributed. In this last case energy conservation strongly constraints the possible outcomes.

Armenise *et al.*[6] considered a complex STS vibrational kinetics for the $CO_2$ molecule, whereby the vibrational modes are not independent, but a reduced model obtained by lowering the dissociation energy was used to decrease the number of vibrational states from 9018 to 1224. In multi-temperature models (e.g. the one by Kustova *et al.*[7]), each vibrational mode is described by a vibrational temperature, but the rapid VV exchange results in the establishment of a Boltzmann distribution with a single temperature $T_{12}$ of the combined symmetric and bending modes.

In plasma conditions, the detailed discussion in Fridman[8] (see also Kozák *et al.*[3]) shows that the most important contribution to dissociation is given by vibrational excitation of the asymmetric mode. These works focus on the kinetics of this mode. In this light, it is assumed that the dominant exchange processes are intra-mode VV where only v changes and corresponding VT and eV processes (e + $CO_2(v) \rightarrow$ e + $CO_2(v')$) involving the asymmetric stretching mode only. To some limited extent the coupling of the asymmetric stretching mode with the other modes has been included in Kozák *et al.*[3] In this work, although we use a conceptually different approach, we are using rate coefficients which are the same as in the aforementioned paper, therefore even in these first calculations, our approach inherits the partial coupling with the other modes.

Even with a single mode, the kinetics is still complex and requires high computational cost, especially when used in 2D or 3D models.

Peerenboom *et al.*[9] applied a dimension reduction method based on principal component analysis to a STS kinetics model of $CO_2$ plasmas. Berthelot *et al.*[10] developed a lumped-levels model to avoid solving equations for all individual $CO_2$ vibrational levels, demonstrating that a 3-groups model is able to (more or less) reproduce the asymmetric mode vibrational distribution function of $CO_2$. Similar procedures where applied in the past to compress the computational requirements of recombining plasma models.[11] In the paper by de la Fuente *et al.*[12], a reduction methodology for the STS kinetics was illustrated, whereby the asymmetric stretching vibrational mode levels are lumped within a single group or fictitious species.

A big help in this effort is provided by analytical results obtained in the past describing specific conditions. For example, under the hypothesis that the i → j and j → i transitions are balanced (i and j are two generic vibrational quantum numbers), and that only intra-mode vibrational exchange processes are important, the Treanor distribution[13]

$$P(v) = A\exp\left(-\hbar\omega v/T_v + x_e\hbar\omega v^2/T_0\right) \qquad (1)$$

is obtained. In the multi-mode case a more general Treanor-Likal'ter[14,15] distribution would be appropriate. In Eq. (1), v is the vibrational level quantum number of the asymmetric mode, $A$ is a normalization factor, $\omega$ is the vibrational quantum in the energy space, $T_v$ and $T_0$ are the vibrational and gas temperature, respectively, $x_e$ is the coefficient of anharmonicity. In $P(v)$, $T_v$ is a parameter related to the internal energy, and describes the population of the first two vibrational levels.

Since the rate coefficients for VT processes increase dramatically for high vibrational quantum numbers leading to a drop in the high energy region of the VDF, previous studies considered these processes as the main limiting factor to achieve effective molecular dissociation.[1,2,4] Since, however, dissociation can be represented as boundary conditions of the kinetic problem, it appears natural to elaborate these concepts using a mathematical approach based on partial differential equations. Accordingly, in this paper we reconsider an alternative approach to STS models, presently overlooked being more mathematically demanding with respect to STS models. This is the approach used in analytical theories developed in the 70's.[4,5,8,16-18]

The approach in question is based on the diffusion approximation[4,16], which transforms the ME for the VDF into a Fokker-Planck (FP) equation. Traditionally, in the FP approach as performed by Rusanov *et al.*[16], equations are solved by assuming a condition of null flux, i.e. $J(\varepsilon) = 0$ where $J$ is the total flux of molecules along the energy axis due to VV and VT processes. Dissociation, of course, is connected to the boundary condition for $\varepsilon = \varepsilon_{diss}$, where $\varepsilon_{diss}$ is the dissociation limit. Therefore one of the important messages of this paper is the fact that the condition of null flux, is not consistent with the presence of the dissociation process: $CO_2$ molecules must experience a net flux from low vibrational levels to the dissociation threshold. An FP approach opens the possibility to include the non-zero flux boundary condition in a logical way.

Our approach is to apply to the FP equation numerical solution techniques in order to avoid the use of approximations which were necessary for the analytical solution; furthermore, we calculate the transport coefficients which enter into the equation from the best available data[3] and provide continuum interpolations consistent with the diffusion method.

**Computational method**

The core of the diffusion approach is the replacement of the system of ordinary differential equations describing the kinetics of discrete levels with a single second order partial differential equation, the FP equation. A large literature is found on the derivation of the FP equation.[4,8,16,19] The use of this equation is justified when transitions between levels close in energy dominate. It implies that molecules are redistributed in energy according to two classes of transport phenomena:

1) the drift, which is deterministic in nature, and moves molecules initially at the same energy at a single new energy after a given time; 2) the diffusion, which is stochastic in nature, and spreads in energy an initial ensemble of particles.

In the diffusion approach, the set of quantum numbers for the vibrational levels is replaced by a continuous energy $\varepsilon$, and two coefficients, a drift coefficient $a(\varepsilon)$ and a diffusion coefficient $b(\varepsilon)$, are introduced instead of the large number of detailed rate coefficients of the STS approach. The drift coefficient measures the speed at which molecules gain (or lose, if negative) energy and therefore is measured in eVs$^{-1}$; the diffusion coefficient is one-half of the rate of increase in variance of the energy distribution of an ensemble of test molecules initially at the same energy, therefore it is measured in eV$^2$s$^{-1}$. These two coefficients, which are functions of $\varepsilon$, are calculated with standard integral formulas[8,19] based on kinetic data. This approach, based on two coefficients only, is accurate for chemical processes where the energy exchange is much smaller than the dissociation energy (the "mean free path" must be smaller than the "gradient scale length" in energy space). Accordingly, usually only mono-quantum processes are included into the description. Although few-quantum transitions can be included, multi-quantum jumps are outside the diffusion approximation. The FP equation is non-linear in the case of $CO_2$ vibrational kinetics due to the nature of the resonant VV process. The problem to be solved has the form:

$$\frac{\partial}{\partial t} f(\varepsilon,t) = \frac{\partial}{\partial \varepsilon}(-af + (b+cf)\frac{\partial}{\partial \varepsilon} f) + R = -\frac{\partial}{\partial \varepsilon} J + R, \qquad (2)$$

where $f(\varepsilon,t)$ is the internal energy distribution of molecules, $a(\varepsilon)$ accounts for the drift due to vibrational excitation by electrons (eV) and VT processes; $b(\varepsilon)$ is the diffusion coefficient due to eV, VT and non-resonant VV processes, $c(\varepsilon)$ is the reduced diffusion coefficient due to resonant

VV processes, $R$ is the sub-threshold dissociation term. Molecules always dissociate when $\varepsilon > \varepsilon_{diss}$, where $\varepsilon_{diss}$ is the dissociation limit, therefore the boundary condition for a solution of Eq. (2) is $f(\varepsilon_{diss}) = 0$.[20] Note the difference between assuming a null value of the distribution and a null value of the flux.

Both $a(\varepsilon)$, the drift coefficient, and $b(\varepsilon)$, the diffusion coefficient, are calculated from the rate coefficients of different chemical processes using the known formulas from the theory of stochastic processes[19]. These rate coefficients are the same used in a STS model (in this work, for example, the ones in the paper by Kozák et al.[3] were used). According to this theory, the diffusion coefficient is given by $b(\varepsilon) \sim \frac{1}{2} d^2 \nu$ where $d$ is the energy exchanged in a single event and $\nu$ is the event frequency per molecule. The drift coefficient $a(\varepsilon)$ is given by $a(\varepsilon) \sim d \nu$. In our case, $d \sim \hbar\omega$, the single process energy exchange calculated by a continuum fit of the energy levels, $\nu$ is the chemical reaction rate per molecule given by $kn'$ where $k$ is the rate coefficient and $n'$ is the number density of the reaction partner. Both the $a(\varepsilon)$ and $b(\varepsilon)$ coefficients are the sum of contributions due to corresponding STS processes. For example, the contribution to the drift of mono-quantum VT processes, $CO_2(v) + M \rightarrow CO_2(v-1) + M$, is given by

$$a_{VT}(\varepsilon) = -(\hbar\omega)k_{VT}(\varepsilon)n_M \qquad (3)$$

and the contribution to the diffusion coefficient $b$ due to VV linear processes ($VV_1$) is given by

$$b_{VV_1}(\varepsilon) = \frac{1}{2}k_{VV_1}(\varepsilon)n_0(\hbar\omega)^2 P_1, \qquad (4)$$

where $n_0$ is the neutral gas density, $k_{VV_1}$ is the rate coefficient for linear VV processes. With respect to the formula (3-127) in Fridman[8], Eq. (4) has been corrected by a factor $P_1$ which is the fractional population of the v = 1 level in a Treanor distribution

$$P_1 = \frac{P(1)}{P(0)} = \exp\left(-\varepsilon_{01}/T_v + x_e \varepsilon_{01}^2/T_0 \hbar\omega\right), \tag{5}$$

where $\varepsilon_{01}$ is the energy difference between v = 0 and v = 1, to account for the fact that the partner of the X(v) + X(1) → X(v+1) + X(0) collision is not any molecule, but only the ones in the v = 1 state. Only collisions with the v = 1 state are included, collisions with higher levels being comparatively less important. The contributions to $a$ and $b$ due to other STS processes are calculated analogously. Since many different values of the rate coefficients may produce nearly the same values of the transport coefficients $a$ and $b$, in this sense a data reduction is achieved.

The transport coefficients are the sum of the contributions of all chemical processes. A fundamental step is based on the principle that, under the hypothesis of null flux, the Treanor distribution (Eq. (1)) must be recovered. This implies that all the processes contributing to $b$, also give a contribution to $a$ according to the formula[8]:

$$a(\varepsilon) = -b(\varepsilon)\left(\frac{1}{T_v} - \frac{2x_e \varepsilon}{T_0 \hbar\omega}\right). \tag{6}$$

The first term into the brackets is easily recognized in terms of detailed balance, or fluctuation-dissipation relation[19], while the second one introduces the effect of the anharmonicity parameter $x_e$ and the gas temperature $T_0$ which may differ from $T_v$ in the non-equilibrium case.

In the continuum formulation it is simple to account for detailed balance. While in the STS approach a list of relations must be satisfied by the VV rate coefficients, in the continuum approach it is sufficient to include an additional term into the expression of the drift coefficient. In this way, following the analytical approach, it can be seen that the Treanor distribution is obtained if the vibrational flux

$$J = af - b\frac{\partial}{\partial \varepsilon} f \tag{7}$$

is set equal to zero (compare with Eq. (1.13) in Rusanov *et al.*[16] and Eq. (13) in Brau[18]). The proposed approach in this paper generalizes that based on assuming $J = 0$. In fact, although this approximation has been useful in the past, it is actually not realistic, since molecules eventually dissociate, and dissociation begins an irreversible diffusion from sub-threshold energies to the threshold energy, implying the presence of a nonzero flux in vibrational energy space. In order to avoid this approximation, the FP equation can be solved by a numerical approach for which the assumption of $J = 0$ is not required.

Several numerical methods are applicable to the solution of the FP equation in the context of our approach: in this paper we use a diffusion Monte Carlo method already applied in the past to similar problems.[21] This method is based on the short-time propagator of the drift-diffusion equation (Eq. (2)) which is given by

$$G(\varepsilon + \xi, h) = \frac{1}{\sqrt{4\pi bh}} e^{-\frac{(\xi - ah)^2}{4bh}}, \tag{8}$$

i.e. the Green function of the Eq. (2) for constant $a(\varepsilon)$ and $b(\varepsilon)$ and neglecting boundary conditions (hence approximated for short time)[22], where $h$ is a numerical time step. The choice of the time step is based on the requirement that the average energy shift is much less than $\varepsilon_{diss}$, say $\varepsilon_{diss}/100$, while a criterion for the convergence of the solution was to check that the steady-state for the function $f$ was reached. The solution of Eq. (2) at time $t+h$ is calculated from the solution at time $t$ using the following convolution equation[22]

$$f(\varepsilon,t+h) = \int G(\varepsilon+\xi,h)f(\varepsilon+\xi,t)d\xi. \qquad (9)$$

Eq. (8) is applied to an ensemble of mathematical dots (walkers) each with a "mathematical weight" $p_i$ and a specified energy $\varepsilon_i$ which represents the VDF when used to estimate macroscopic quantities.

Basically any walker at any time step performs two moves, the first one deterministic and the second one stochastic. The first move takes into account drift described by the $a$ coefficient and it is simply an energy shift given by $+a(\varepsilon)h$. The second move produces a random shift according to the Gaussian distribution of variance $2b(\varepsilon)h$.

This method allows a simple treatment of boundary conditions and is free from numerical diffusion, which means that a Gaussian solution propagates under a constant $a(\varepsilon)$ and null $b(\varepsilon)$ in the exact way, by shifting its center of mass. Since the solution ranges several orders of magnitudes, a variance reduction, based on the Russian roulette method,[23] is implemented to reduce the computational cost.

Non-linear VV processes (VV$_n$), e.g. $2X(v) \rightarrow X(v-1) + X(v+1)$ are included in a straigthforward way. The reduced diffusion coefficient $c$ in Eq. (2) is given by $(1/2)k_{v,v-1}^{v,v+1}(\hbar\omega)^2$

. In a time-dependent approach, $f$ is known from previous calculations, and using Eq. (2) $c$ is calculated and summed as a non-linear contribution to the coefficient $b$ in Eq. (7).

Sub-threshold dissociation processes are described by the term $R$ in Eq. (2). This term has the form $R = -k'n'$, where $k'$ is the corresponding rate coefficient, and has dimensions [1/time]. Therefore sub-threshold dissociation is included by a removal process of the walker, with probability $p = 1-exp(-Rh)$. For both this process and the threshold process, the removed walkers are redistributed randomly in order to keep the normalization constant.

The function $f$ must be normalized appropriately if STS rate coefficients are used to calculate the non-linear diffusion coefficient $c$. This is also useful in order to compare the results of the diffusion approach to STS calculations. The STS populations are attributed to levels which are limited in number and the populations of such levels sum to $n_0$, the total number density of molecules, while the normalization of $f$ is based on its integral in the $(0, \varepsilon_{diss})$ range. An appropriate normalization can be formulated based on the equilibrium case where the population of each STS level is given by the Boltzmann distribution, therefore its value for v = 0 is given by $n_0/Z$, $Z$ being the partition function. The corresponding continuous function has a value of $n_0/T_v$ (the energy is defined in such a way that the energy for the v = 0 level is zero): This means that all energies are scaled down by the zero-point energy of the considered degree of freedom. Therefore, the appropriate normalization is established by setting

$$\int_0^{\varepsilon_{diss}} f(\varepsilon)d\varepsilon = n_0 \frac{T_v}{Z}. \tag{10}$$

In such a way, the level populations are matched in the equilibrium case. In this equation, $Z$ (expressed in m$^{-3}$) is actually a cut-off partition function, i.e. the sum of the populations of the very first levels is a Boltzmann distribution at the temperature $T_v$. The exact number of levels > 2 is immaterial at normal values of $T_v$, and the difference between this sum and the same calculated using a Treanor distribution is very small. The full function based on the Treanor distribution is not an appropriate choice since the Treanor distribution increases exponentially at high energy.

In order to calculate the *a* and *b* coefficients and the *c* reduced diffusion coefficient due to $VV_n$ processes in the $CO_2$ molecule case, a continuous polynomial fit of the $VV_1$, $VV_n$ and VT rate coefficients at 300 K of the data in Figure 2 in Kozák *et al.*[3] as a function of $\varepsilon$ (not of the discrete vibrational level index v) is included, as well as a fit of the average energy exchange $\delta\varepsilon(\varepsilon)$ of these processes. For the $VV_1$ and $VV_n$ processes $x_e$ in eq. (6) was set equal to $5.25 \times 10^{-3}$, as in reaction (V8) in Kozák *et al.*[3] A functional expression of the rate coefficients $k(\varepsilon)$ as functions of $\varepsilon$ is necessary in some expressions for *a* and *b* as seen above. To this aim, we have interpolated the corresponding STS rate coefficients after replacing the quantum number v with the corresponding energy as independent variable. The interpolations used in this work are for a fixed $T_0$ and *log k* or *k*, depending on the case, is fitted as $\sum_{i=0}^{4} c_i \varepsilon^i$. To calculate the rate coefficients at different temperatures, the temperature dependencies reported by Kozák *et al.*[3] are used. Note that VV'$_a$ and VV'$_b$ processes in Kozák *et al.*[3] are here considered as VT processes, as also recommended in Kozák *et al.*[3] Those processes are VV' relaxations between the asymmetric and the first two symmetric mode levels, $CO_2(v) + CO_2 \rightarrow CO_2(v-1) + CO_2(v_a)$, $CO_2(v) + CO_2 \rightarrow CO_2(v-1) + CO_2(v_b)$.

As sub-threshold dissociation process we have included reaction (N1) in Kozák *et al.*[3], that is $CO_2(v) + M \rightarrow CO + O + M$, where M is any neutral species, assuming for simplicity collisions with CO$_2$ molecules (in any vibrational state) only. The processes included in the present calculations are summarized in Table 1.

**Results and Discussion**

In Figure 1, the steady-state solution of Eq. (2) is reported for $n_0 = 2.33 \times 10^{23}$ m$^{-3}$, $T_0 = 300$ K and three different values of the parameter $T_v$. The value for the CO$_2$ number density is based on the results reported on Figure 7 (the 8 ms case) in Kozák et al.[3] for a power density of 30 W cm$^{-3}$ and a pressure of 2660 Pa which are typical for a CO$_2$ conversion reactor.[24,25]

As can be seen, the solution for high energies is strongly sensitive to the value of $T_v$. For comparison, the result of STS calculations by Kozák et al.[3], based on the same values of the rate coefficients, is reported. The scheme is able to capture the trend of the STS calculations which correspond to a vibrational temperature of 0.19 eV (an estimate based on using the Boltzmann distribution then slightly higher than 0.18 eV here) and therefore is also semi-quantitatively compatible. It is not necessary to match $T_v$ exactly, since this last has a different definition in the discrete and continuum regime. In fact, $T_v$ here is a parameter inside the transport equation, whereas in the STS approach it is deduced *a posteriori* from the $n(v = 1)/n(v = 0)$ population ratio. It should be noted that we do not include linear VV collisions with states different than v = 1, differently from what is done in Kozák et al.[3] This proves that v = 1 is the most important level and that with the inclusion of those collisions only, most of the VDF and all the essential features obtained with the STS model can be reproduced. Also, we do not include reaction (N2) in Kozák et al.[3], that is CO$_2$(v) + O → CO + O$_2$, since there was no information available on the number density of O atoms. This could affect the shape of the VDF in some conditions, as shown in Berthelot and Bogaerts.[26] However, apparently it is not very important in this case, since we almost obtain the same VDF.

These calculations allow to perform a first quantitative discussion of the actual transport processes of molecules along the vibrational energy scale. In particular, it is possible to characterize the drift and diffusion coefficients for different vibrational and gas temperatures and to calculate the contribution of the chemical processes to the energy flux $J$ which is connected to molecule dissociation.

Indeed, as shown in Figure 2, our formulation shows the role of the real main factors determining the shape of the VDF. Including only $VV_1$ processes with appropriate boundary conditions, the familiar non-equilibrium shape with a long plateau (lower curve) is obtained. Only when non physical reflecting boundary conditions (i.e. $J = 0$) at the dissociation threshold are used (upper curve) the Treanor distribution appears. This also demonstrates that detailed balance is not a sufficient condition to obtain the Treanor distribution, that appears only when the requirement of a reflecting boundary condition is selected. With the assumption of reflecting boundary conditions, VT processes can play a role and display a qualitatively satisfactory trend (intermediate curve), but this trend is still far from the solution of the complete equation, and VT processes play a negligible role in the conditions of the present calculations when the appropriate boundary is used. Under the null flux hypothesis, VT processes are found to be essential to retrieve the characteristic shape (including a high slope bulk, a low slope plateau, and high slope tail) of the vibrational distribution.[16] The same concept applies to STS as well, in the sense that a Treanor distribution must result from STS calculations if the dissociation process and VT processes are removed. In Figure 7 in Berthelot and Bogaerts,[26] consistently with our general theory, a raise of the VDF is observed removing the dissociation process in STS calculations, although they do not obtain an actual Treanor distribution. These findings are discussed more quantitatively in the next figures.

In particular, in Figure 3, the coefficients $a$ and $b$ for the distribution with $VV_1$ processes only, that reproduces the Treanor distribution in Figure 2, are reported. It can be seen that the $b$ coefficient has a relatively weak dependence on the energy, while the drift coefficient $a$ increases steadily and changes sign at a defined energy depending on the value of the $T_v/T_0$ parameter. This feature is mostly an effect of the second term inside parentheses of Eq. (6) for $a$ and therefore it is a result of the anharmonicity of the oscillators. This sign change determines most of the shape of the VDF as shown above.

In Figure 4, the coefficients $a$ and $b$ for the case in Figure 1 that reproduces closely the results in Kozák et al.[3] ($T_v = 0.18$ eV) are displayed. This figure shows that the diffusion approach produces an effective alternative interpretation of the main phenomena occurring in the vibrational kinetics. The VDF appears as the result of the diffusion in energy space due to VV processes, accelerated in the middle-v region by the drift due to linear VV process. VT processes never play a significant role, their contribution to drift being much lower than VV and VV' processes. It can be seen that the non-linear VV processes are significant only in the low energy region, not an important issue since the low energy part is close to the Treanor distribution.

In Figure 5 steady-state results are shown for a higher vibrational temperature of 0.25 eV ($n_0$ and $T_0$ were not changed with respect to the previous cases). In Figures 6 and 7 the transport coefficients for two of the curves in Figure 5 are shown. In this case, the energy corresponding to the minimum of the Treanor distribution is lower. The plateau of the VDF begins correspondingly at a lower energy. Figure 6, when compared to Figure 3, shows that the main effect of increasing $T_v$ is the shift of the drift coefficient $a$ curve to lower energies while $b$ is not much affected. Again, the second term in Eq. (6) makes the difference.

To better demonstrate this point, Figures 8-9 show the coefficients $a$ and $b$ for different values of the gas and vibrational temperature. These plots show that the "no return" energy level beyond which a molecule most likely dissociates shifts to lower energies when $T_0$ is reduced. We believe that this is the main reason of the higher energy efficiency found in low $T_0$ reactors, like the ones exploiting plasma vortexes and expansions[1]. In particular, the main parameter appears to be the temperature ratio $T_v/T_0$ which is related to the energy of the Treanor minimum. Of course, in cases like the ones in Figures 8-9 where the $a(\varepsilon)$ coefficient is always negative, a "no return point" is never reached, but molecules can dissociate anyway due to the random processes of diffusion: the dissociation rate is correspondingly low in such cases.

As a consequence of the no-return energy concept, VT processes can hardly affect dissociation even under conditions where they could affect the VDF. In fact, high energy molecules are pushed by the drift toward dissociation. VT processes may push molecules backwards, but this will only produce an increase of the VDF to compensate the effect. This situation can change under high temperature conditions where VT processes may become important for vibrational levels below the Treanor minimum.

While these results show the critical role played by the temperature ratio $T_v/T_0$ as a parameter, it is true that the value of this parameter must be determined (if the case, as a function of time) in a full model. This has been done in recent STS models by including eV processes and their corresponding exothermic, second-kind collisions.[3,26-28] However, only the very first levels enter the energy balance which determines $T_v$ for a given $T_0$, therefore the diffusion model could be integrated in the next future by a very reduced STS energy balance involving only these levels, e.g. three of them. A simple energy balance able to estimate the value of $T_v$ can be established based on the populations of the lowest vibrational levels, even just v = 0 and v = 1. The balance is based on the eV processes (v = 0) → (v = 1) and (v = 1) → (v = 0) and the VV$_1$ process (v = 1) + (v = 1) → (v = 2) + (v = 0). A simplified steady-state population balance is written in the form:

$$k_{eV01} n_e n(0) = k_{eV10} n_e n(1) + k_{10}^{12} n(1)^2$$
$$n(0) = \frac{n_0}{Z}, \quad n(1) = \frac{n_0}{Z} \exp(-\varepsilon_{01}/T_v + x_e \varepsilon_{01}^2 / T_0 \hbar \omega) \quad (11)$$

Note that Eq. (11) differs from the usual form of the simplified population balance in using the Treanor instead of the Boltzmann distribution. In view of results of the present study (e.g. Figure 1) this is expected to be generally more accurate. Furthermore, in the diffusion approach, $T_v$ is a parameter describing the low energy behaviour of the solution of Eq. (2), therefore, Eq. (1) fixes it unambiguously even for strong non-equilibrium cases. Since, with good approximation, $k_{eV01}$ and $k_{eV10}$ are related by the detailed balance relation (exact in the case of a Maxwellian EEDF), Eq. (11) is readily solved producing $T_v$ as a function of $T_e$ and $n_e$. In the case of strongly non-equilibrium EEDFs, the usual approach based on the two-term Boltzmann equation[2, 27] can be used to calculate $k_{eV01}$ and $k_{eV10}$. Eq. (11), therefore, implies that the most important processes controlling the diffusion flux $J$ are the (v = 0) → (v = 1) eV transition and the (v = 1) + (v = 1) → (v = 0) + (v = 2) VV process. Furthermore, from the opposite point of view, the diffusion approach could be used to extrapolate, and accelerate considerably, the STS model from the lowest levels upwards. This perspective is very promising for future STS models including detailed coupling to the other two vibrational modes, or even rotational levels.

**Conclusions**

In this work the diffusion approach was used to study the vibrational kinetics of $CO_2$ molecules in the context of plasma dissociation. To this aim, the FP equation is solved numerically in order to avoid the use of strong approximations.

Explicit formulas for the transport coefficients *a*, *b* and *c*, this last describing the non-linear effects, are obtained based on the theory of stochastic processes and interpolation of STS rate coefficients. Results are found to be in good agreement with STS calculations in the literature. The results assuming the null flux approximation are reproduced by preventing molecules to dissociate even when reaching $\varepsilon_{diss}$ (i. e. a reflecting boundary condition), which, of course, is not physical. Under such conditions, the Treanor distribution is found for a long enough time, a result fully consistent with Treanor's original derivation.[13] By removing this condition and allowing molecules to dissociate, the plateau in the VDF is obtained even using only linear VV processes. This demonstrates that the essential features of the VDF are mostly a result of the dissociation process which acts like a boundary condition, removing the exceedingly restrictive null flux condition. Our results show that $CO_2$ molecules reach dissociation threshold under the effect of a positive vibrational drift which onsets at the energy corresponding to the minimum of the Treanor distribution obtained in the null flux case. This conclusion implies that $CO_2$ molecules are most likely to dissociate after reaching the no-return energy point which becomes lower when the gas temperature is decreased. This effect, more than the temperature dependence of VT processes, may explain why low gas temperature non-equilibrium plasmas are producing such good results in terms of obtaining high energy efficiency to dissociatie molecules. Of course this finding is specific to $CO_2$, but in perspective our approach can be applied to the dissociation kinetics of other molecules like $CH_4$ or more complex ones, allowing to determine the role of boundary conditions and the main factors affecting molecular dissociations in such cases.

In this respect, our method requires, in order to calculate appropriate transport coefficients, the determination of very accurate rate coefficients sets specially in terms of consistency of trends (with quantum numbers, with temperature). New rate coefficients sets, furthermore, for the three normal modes, for example involving $CO_2(i, j, k)$ become timely and usable in view of the possibility of a multi-dimensional generalization of the continuum approach.

**Acknowledgements**

**Table 1.** Elementary reactions used in the calculations. All rate coefficients are calculated using data in Kozák *et al.*[3]

| Name | Reaction | Note |
|---|---|---|
| $VV_1$ | $CO_2(1) + CO_2(v) \rightarrow CO_2(0) + CO_2(v+1)$ | |
| $VV_n$ | $CO_2(v) + CO_2(v) \rightarrow CO_2(v-1) + CO_2(v+1)$ | |
| VT | $CO_2(v) + CO_2 \rightarrow CO_2(v-1) + CO_2$ | a |
| Dissociation | $CO_2(v) + CO_2 \rightarrow CO + O + CO_2$ | b |

[a] Sum $VT_a + VT_b + VT_c + VV'_a + VV'_b$ in Kozák *et al.*[3]
[b] Sub-threshold contribution

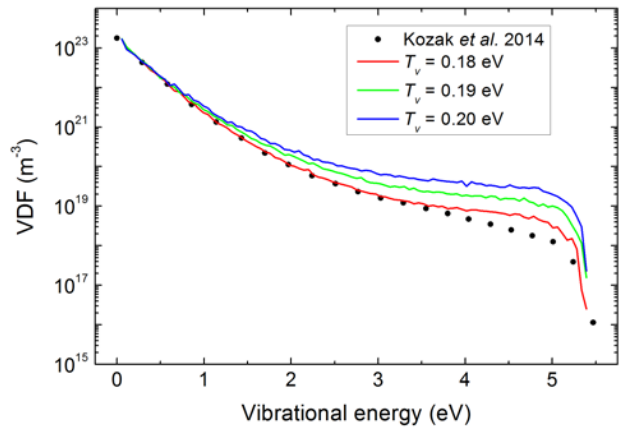

**Figure 1.** Demonstration that the improved FP approach catches the essential features of the Master Equation. Here the vibrational distribution function of the asymmetric mode levels of the $CO_2$ molecule is calculated for different values of the vibrational temperature. Results based on expressions of *a*, *b, c* and *R* obtained from rate coefficients in Kozák *et al.*[3] Dots represent STS calculations results in Figure 7 (8 ms) in Kozák *et al.*[3] for a power density of 30 W cm$^{-3}$ and a $T_0$ of 300 K, giving a $T_v$ of 0.19 eV, determined from the ratio of the populations in the v = 0 and v = 1 levels.

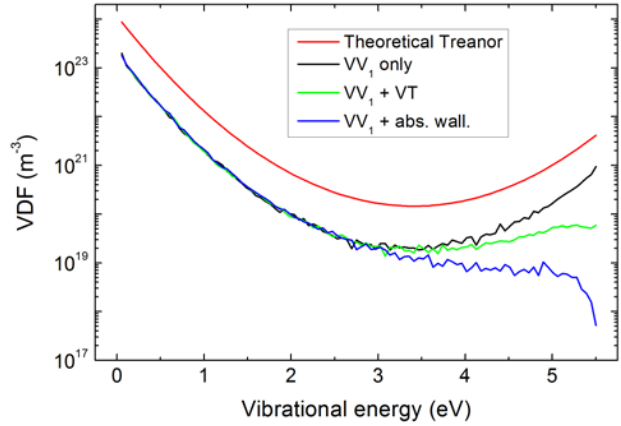

**Figure 2.** Vibrational distribution functions for $n_0 = 2.33 \times 10^{23}$ m$^{-3}$, $T_0 = 300$ K and $T_v = 0.19$ eV, obtained for different choices of the boundary condition and processes: VV$_1$ processes only (black line), VV$_1$ and VT processes (green line) and absorbing wall boundary condition (blue line). The theoretical Treanor distribution function (red line, shifted upwards for better representation purposes) is also shown.

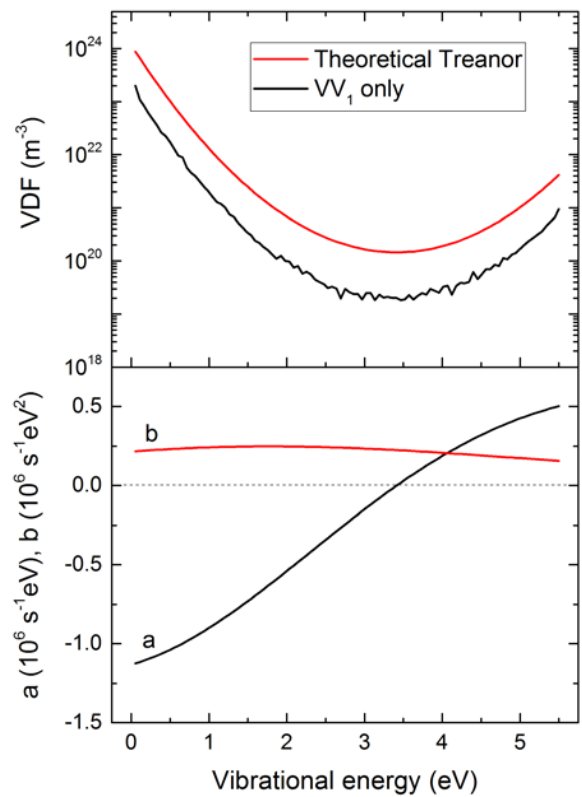

**Figure 3.** Vibrational distribution function with $VV_1$ processes only, that reproduces the Treanor distribution for $T_v = 0.19$ eV and $T_0 = 300$ K (see Figure 2) (top) and corresponding coefficients *a* and *b* (bottom). In the top panel, the theoretical Treanor distribution function (shifted upwards for better representation purposes) is also shown.

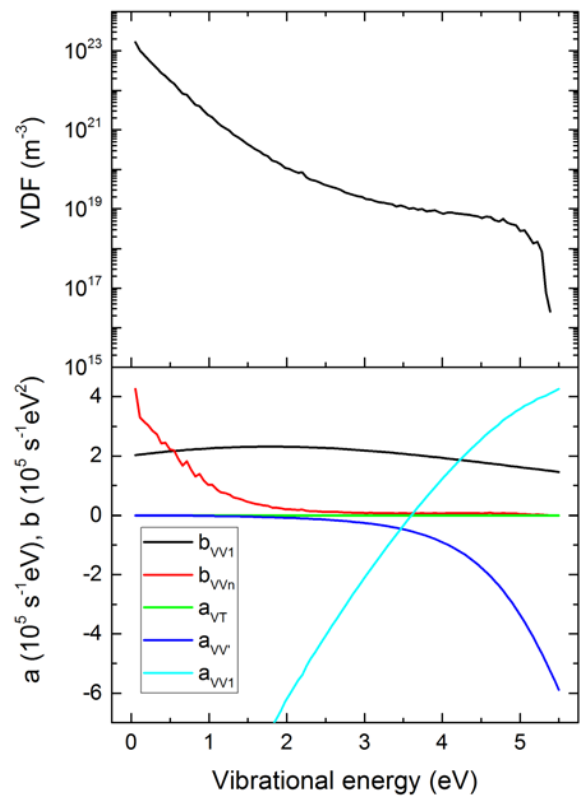

**Figure 4.** Vibrational distribution function for the case closer to the results in Kozák *et al.*[3] ($T_v = $ 0.18 eV and $T_0 = $ 300 K, see Figure 1) (top) and corresponding coefficients *a* and *b* (bottom).

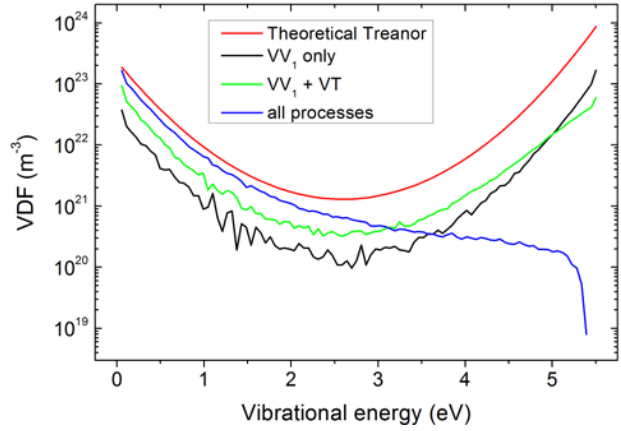

**Figure 5.** Vibrational distribution functions for $n_0 = 2.33 \times 10^{23}$ m$^{-3}$, $T_0 = 300$ K and $T_v = 0.25$ eV, obtained for different choices of the boundary condition and processes. The theoretical Treanor distribution function (shifted upwards for better representation purposes) is also shown.

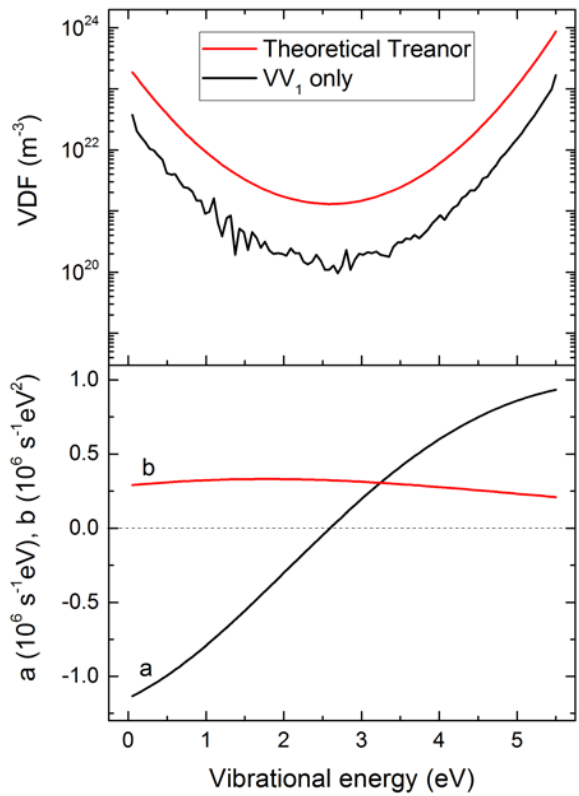

**Figure 6.** Vibrational distribution function with $VV_1$ processes only, that reproduces the Treanor distribution for $T_v$ = 0.19 eV and a $T_0$ = 300 K (see Figure 5) (top) and corresponding coefficients *a* and *b* (bottom). In the top panel, the theoretical Treanor distribution function (shifted upwards for better representation purposes) is also shown.

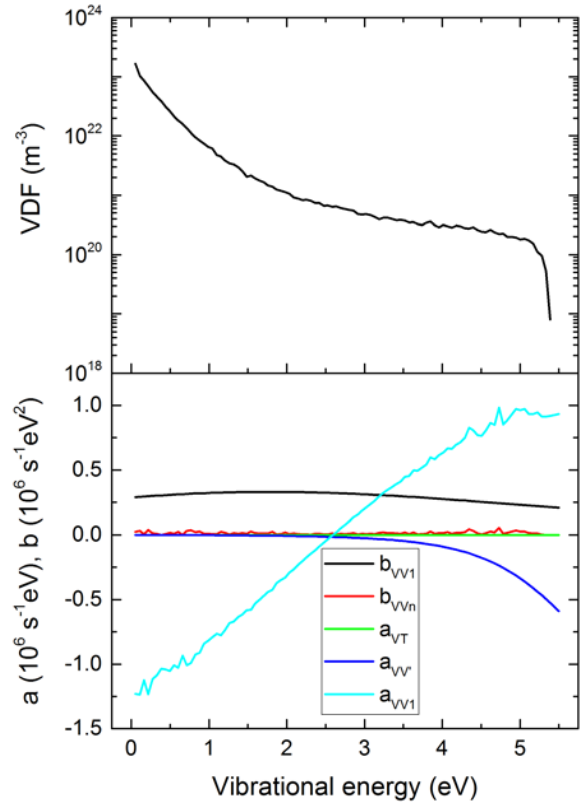

**Figure 7.** Vibrational distribution function with all the processes included ($T_v = 0.25$ eV and a $T_0 = 300$ K, see Figure 5) (top) and corresponding coefficients *a* and *b* (bottom).

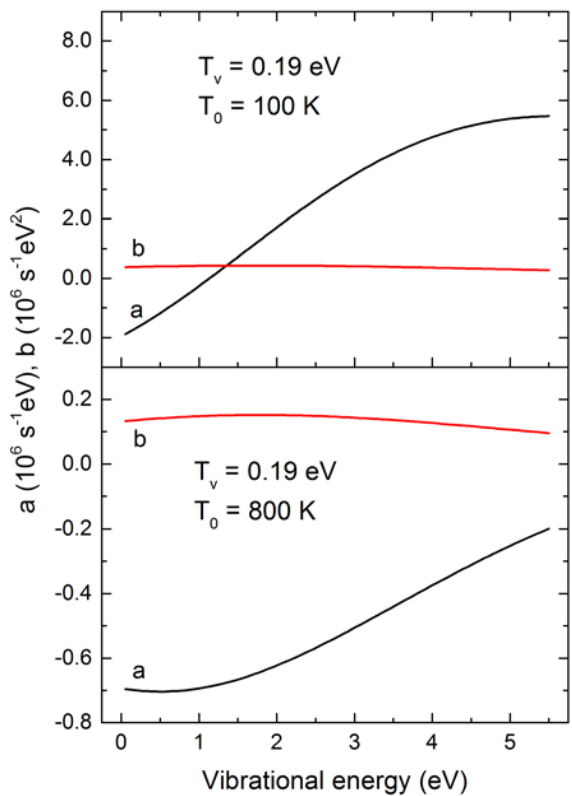

**Figure 8.** Coefficients *a* and *b* for the distribution with VV$_1$ processes only, for $T_v = 0.19$ eV and $T_0 = 100$ K (top) and 800 K (bottom).

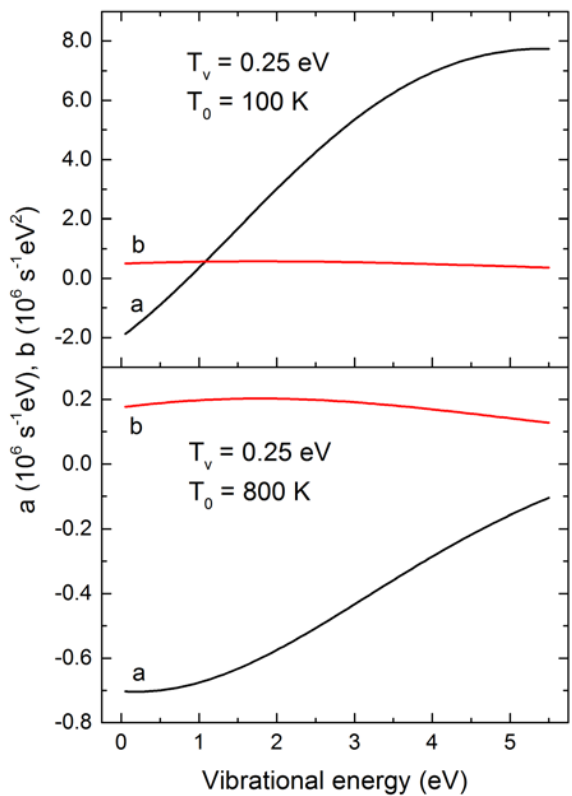

**Figure 9.** Coefficients *a* and *b* for the distribution with $VV_1$ processes only, for $T_v = 0.25$ eV and $T_0 = 100$ K (top) and 800 K (bottom).

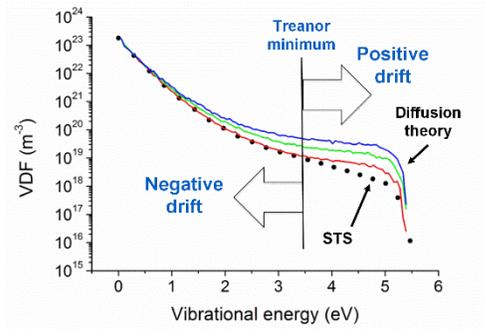

**TOC Graphic**